\documentclass[proof]{pasj01}

\usepackage{natbib}

\usepackage{mathtools} 

\usepackage{lscape} 
\begin{document}

\Received{}
\Accepted{}


\title{Development of cross-correlation spectrometry \\and the coherent structures of maser sources}


\author{Kazuhiro \textsc{Takefuji}\altaffilmark{1}}
\email{takefuji@nict.go.jp}

\author{Hiroshi \textsc{Imai}\altaffilmark{2}}

\author{Mamoru \textsc{Sekido}\altaffilmark{1}}

\altaffiltext{1}{National Institute of Information and Communications
Technology, 893-1 Hirai, Kashima, Ibaraki 314-8501, Japan} 
\altaffiltext{2}{Science and Engineering Area of
the Research and Education Assembly, Kagoshima University, 1-21-35 Korimoto, Kagoshima
890-0065, Japan}


\maketitle

\def\h2o{H$_2$O}
\def\kms{~km~s$^{-1}$}

\begin{abstract}
We have developed a new method of data processing for radio telescope observation data to measure time-dependent temporal coherence, and we named it cross-correlation spectrometry (XCS).
 XCS is an autocorrelation procedure that expands time lags over the integration time and is applied to data obtained from a single-dish observation. 
 The temporal coherence property of received signals is enhanced by XCS.
 We tested the XCS technique using the data of strong H$_{2}$O masers in W3 (H$_2$O), W49N and W75N. We obtained the temporal coherent lengths of the maser emission to be 17.95 $\pm$ 0.33 $\mu$s, 26.89 $\pm$ 0.49 $\mu$s and 15.95 $\pm$ 0.46 $\mu$s for W3 (H$_2$O), W49N and W75N, respectively. These results may indicate the existence of a coherent astrophysical maser.
  
\end{abstract}

\section{Introduction}
Historically, the Michelson interferometer (\citealt{1887SidM....6..306M}) was first used to measure temporal coherence of light. Light input to the Michelson interferometer is split into two paths by a half-mirror. Then the split light beams are combined after reflection by total-reflection mirrors to form interference fringes. The temporal coherence can be measured by changing the position of the total-reflection mirrors, where the correlation is taken between the light signals originating from the same source but arriving at a detector slightly at different epochs. 

We consider an introduction of this technique to measurement of the temporal coherence of an astronomical radio signal using a radio telescope as follows. The observed data is recorded digitally in hard disks, then the temporal coherence of the observed signal is measured by autocorrelation while applying time lags over the integration time. Instead of taking the simple autocorrelation, a single time-series of data is divided into small chunks corresponding to short time slots, and each chunk of data is converted to a frequency spectrum by Fourier transform. The time-series of the complex frequency spectrum data is then used for coherence analysis. We named this algorithm  cross-correlation spectrometry (XCS).

This paper describes the XCS algorithm and examples of its application to exploration of the coherence of interstellar water (H$_2$O) maser emission. \h2o\ masers are important tools in astrophysics and astrometry in terms of their limited association with specific evolutionary stages or physical conditions of astronomical objects and their extreme brightness and compactness (e.g., \citealt{elitzur}; \citealt{Gray2012}). Astronomical \h2o\ masers have been considered to generate ``incoherent masers'', which produce electromagnetic waves originating from different volumes of the maser region and with random wave phases. However, in an extreme case, where the maser emission is generated from a much smaller region ($l<<$0.1~AU) in a much sharper beam ($\theta <<$0.01~rad) and at much higher brightness temperature ($T_{\rm b}>> 10^{14}$~K),  ``coherent maser'' emission is expected, in which specific electromagnetic waves with specific synchronized phases are greatly enhanced (e.g., \citealt{1991SvAL...17..250I}; \citealt{elitzur}). 
Without extremely high angular resolution, it is impossible to spatially resolve such coherent emission regions in the observed astronomical masers. We demonstrate that the XCS technique can be used to distinguish coherent emission even in single-dish observations. 
 
\section{Development of Cross-Correlation Spectrometry (XCS) }
In digital FX-type spectrometers, Fourier transform and
multiplication are applied to data to
obtain a power spectrum of the signal (\citealt{thompson}). 
Here we assume that the spectrum of data obtained with a receiver output of a
radio telescope $X(f)$ is composed of a signal $S(f)$ arriving from a radio
source in the sky and the system noise $N(f)$. The signal of $X(f)$ is dominated by
the receiver and off-source sky noise in most cases.
 The power spectrum is obtained by calculating the inner product of
$X(f)$ and  its complex conjugate $X^*(f)$ as follows: 
\begin{equation}
\begin{split}
|X (f)|^2 & =X (f)X^{*} (f)  \\ 
             &=S (f)S^*(f)+S(f) N^*(f)+S^*(f) N(f)+N(f) N^*(f). \label{eqn1}
\end{split}
\end{equation}
	
The second and third terms of the right-hand side in equation \ref{eqn1} are  
products of the signal and noise. Since the signal and noise are independent,
these cross terms vanish upon averaging. Consequently,
the signal and noise terms remain to give
\begin{equation}
<|X(f)|^2> \cong <S(f) S^*(f)> + <N(f) N^*(f)>, 
\end{equation}
where the bracket indicates averaging over the data accumulation period.

\begin{figure}[htbp]\centering 
\includegraphics[scale=0.7,angle=0,trim=0 0 0 0]{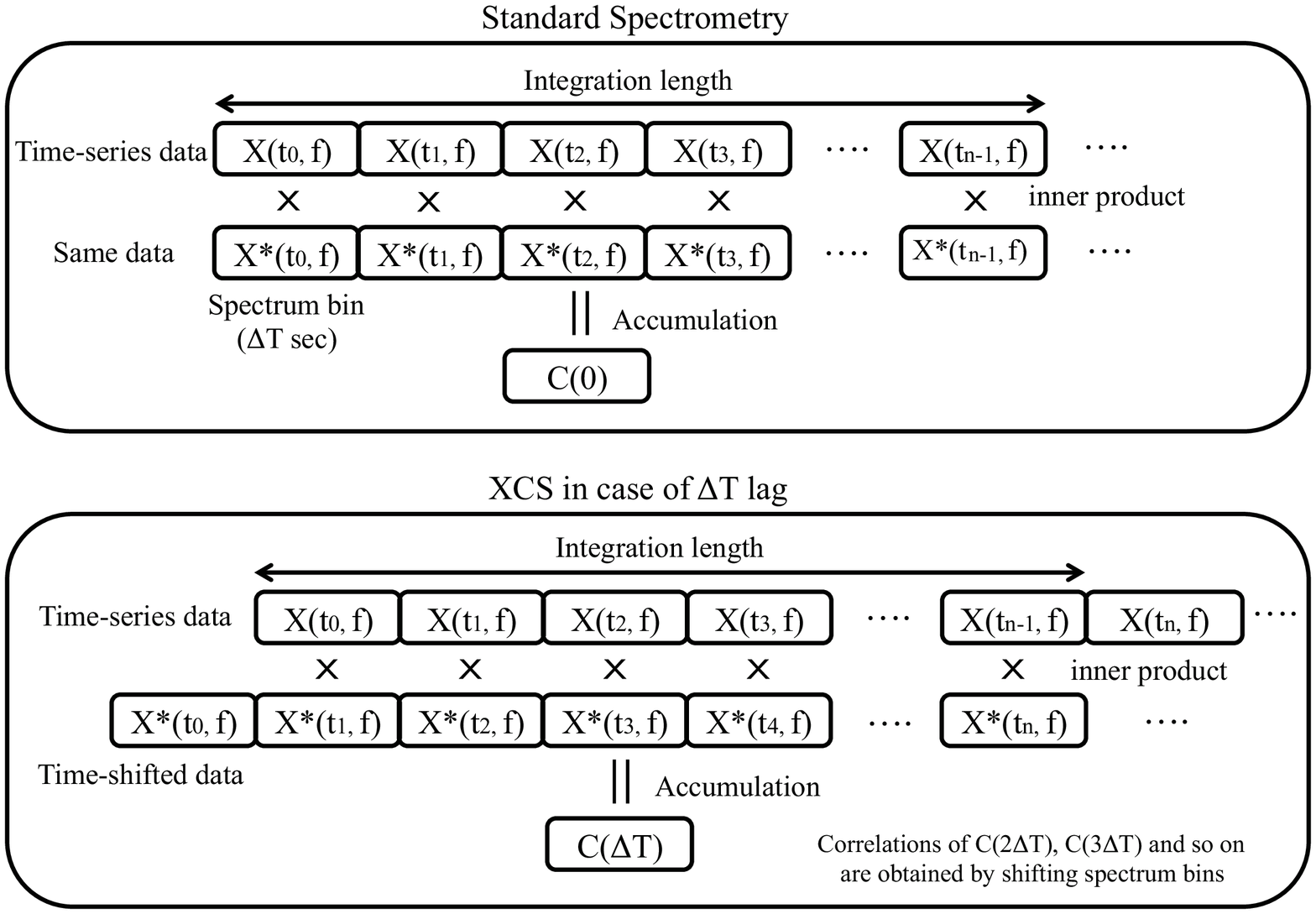}
\caption{Schematic diagrams of standard spectrometry and XCS. We first perform a Fourier transform to observed time-series data with Fourier length $\Delta T$ before both methods. The main difference between the two methods is that data are time-shifted in XCS.}
\label{fig:concept} 
\end{figure}

Next, we describe the algorithm of XCS. A series of data digitized by a sampler is divided
into datasets with time interval $\Delta T$. The frequency spectrum of each dataset is computed by Fourier transform. 
Here $X(t_{ i},f_{ k})$ indicates the complex component at frequency
$f_{ k} = k / \Delta T $ and epoch $t_{ i} = i \times \Delta T$ ($i=0,1,2, ...$ and $k= 0,1,2, ...$), and is expressed as
\begin{align}
X(t_{ i}, f_{ k}) &= S(t_{ i}, f_{ k})
 +  N (t_{ i}, f_{ k}). 
\end{align}
Computing the time-delayed correlation of the data along 
time series $t_{ i}$ at frequency $f_{ k}$, the averaged 
cross terms of the signal and noise vanish because of their independence. Noted that the fourth term of the right-hand side in equation \ref{eqn2} also vanishes because of both noises are taken in different epochs and therefore independent. 
When the signal has temporal coherence, only the first term will remain:
\begin{align}
 < X(t_{ i}, f_{ k}) X^*(t_{ i+j},f_{ k})> &=
 <S(t_{ i},f_{ k}) S^*(t_{i+j}, f_{ k})>
 + <S(t_{ i}, f_{ k}) N^*(t_{ i+j},f_{ k})>   \notag \\
 & + <N(t_{ i},f_{ k}) S^*(t_{ i+j}, f_{ k})>
 + <N^* (t_{ i},f_{ k}) N^*(t_{ i+j},f_{ k})>  \label{eqn2} \\
 &\cong <S(t_{ i}, f_{ k}) S^*(t_{ i+j},f_{ k})>,
\end{align}
where $j$ is the lag number ($j \geq 1$) in the correlation function.
By computing this quantity, the temporal coherence of the signal is measured. Figure \ref{fig:concept} shows schematic diagrams of standard spectrometry and XCS. The main difference between the two methods is that data are time-shifted in XCS.  The minimum time-shift, $\Delta$T, is equivalent to one Fourier length in the figure in XCS side and an important parameter for the XCS algorithm. We called this time shift "forbidden time lag".  Within the time lag, the internal random noise has correlation and unwanted signals like system and sky noises will not be suppressed. Consequently, bandpass profile similar to that of normal spectroscopy will be formed. Therefore, calculation with a delay shorter than it should be avoided for the coherence calculation.

On the basis of the above discussion, we developed a program to perform XCS by modifying cross-correlation software.
Here we describe the details of the XCS algorithm. 
A received signal $x(t)$ is expressed as a sum of Fourier components at frequencies $f_k (k=0,1,2,\cdots,M)$, whose amplitude and initial phase
are $A_k$ and $\theta_k$, respectively, i.e., 
\begin{equation}
\displaystyle{}  x(t) = \sum_{ k=0}^{M-1} A_k \exp \{ 
 j 2 \pi f_{ k} t + j \theta_k \}.  
\end{equation}
This signal is expressed in frequency domain as
\begin{equation}
\displaystyle{}  
 X(f) = \sum_{ k=0}^{M-1} A_k \delta( f - f_k) 
 \exp \{j \theta_k \}.
\end{equation}
Each frequency component is expressed using Kronecker's delta function. Here we
assume  that a series of data is divided into multiple datasets with  time
span $\Delta T  = M \Delta t$, then 
 the frequency resolution in the discrete Fourier transform 
is $\Delta f = 1/ \Delta T$ Hz,
where $\Delta t$ is the time interval used in digital sampling or the maximal sampling speed.
The data sample at epoch $l \Delta t$ in dataset $i$ ($t= i \times \Delta T + l \times \Delta t $) is expressed as 
\begin{equation}
 \displaystyle{} 
 x_i( \Delta t l) = \sum_{ k=0}^{M-1} 
 A_{k,i} \exp \{j [(2 \pi f_{k}  (i \Delta T  + l \Delta t  ) + \theta_{k,i}] \},
\end{equation}
 where $f_{ k} = \Delta f \times k$, $l=0,1,2,\cdots M-1$, and
$\theta_{k,i}$ denotes the phase of spectrum $k$ at the beginning epoch of dataset $i$. After the Fourier transform,
the spectrum component of  dataset $i$ becomes
\begin{equation}
\displaystyle{} 
X_{i}(t_{i}, f) = \sum_{ k=0}^{M-1} A_{k,i}
\delta(f - \Delta f k)   \exp \{j  \theta_{k,i} \}. \label{eqn:f}
\end{equation}
Here we consider a single Fourier component at frequency $f_k$,
\begin{equation}
 X_{i}(t_{i}, f_{k}) = A_{k,i} \exp \{ j \theta_{k,i} \}.
\end{equation}
The time-delayed autocorrelation  $C_k (n \Delta T)$ is expressed
  as a product of the correlation coefficient and its complex
conjugate with time lag $n \Delta T $ as  
\begin{equation}
 C_{k}(n \Delta T) = < X_i (t_{i}, f_{k}) X_{i+n}^*(t_{i+n},f_{k}) >.
\end{equation}
 We assume that XCS will be performed on a time scale of microseconds. If the amplitude of the signal
 $A_{k,i}$ changes slowly on a much longer time scale, the ensemble averages of the amplitude and phase can be separated in the derivation of the following equation:
\begin{align}
C_{k}(n \Delta T) 
 = < A_{k,i} A^*_{k,i+n} \exp \{ j \delta \theta_{k,n} \} > \label{eqn:Cff1}\\
 \cong < A_{k,i} A^*_{k,i+n}> <\exp \{ j \delta \theta_{k,n} \}>. \label{eqn:cff2}	
\end{align}
The temporal coherence of the signal is expressed by a temporal average of $\exp \{ j \delta \theta_{k,n} \}$,
where $\delta \theta_{k,n} \equiv \theta_{k,i}-\theta_{k,i + n}$ and
 $\delta \theta_{k,n}$ is the phase difference
at an interval of $n\Delta T$. Hence, the temporal coherence coefficient is obtained as  
\begin{equation}
 <\exp \{ j \delta \theta_{k,n} \}> =\frac{C_k(n \Delta T)}
  {< A_{k,i} A^*_{k,i+n}>}. \label{eqn:cohrence-cff}
\end{equation}
Equation (\ref{eqn:cohrence-cff}) has a form similar to the coherence spectrum (\citealt{white}, \citealt{1972ITAP...20...10I}, \citealt{bry2014}). The coherence spectrum $coh(f,t)$ of two signals $X$ and $Y$ at frequency $f$ and time $t$ is expressed as
\begin{equation}
 coh(f,t) =\frac{|C_{XY}(f,t)|}{\sqrt{C_{XX}(f,t)}\sqrt{C_{YY}(f,t)}}, \label{eqn:cs}
\end{equation}
where $C_{XX}(f,t)$ and $C_{YY}(f,t)$ are the autocorrelation spectra of the two signals and $C_{XY}(f,t)$ is the cross spectrum of the two signals.  Since our main target is to measure the coherence making the time-shift, XCS can be considered as a special application of the coherence spectrum. 
\section { Observations \label{sub:xcs}} 
\begin{figure}[htbp]\centering 
\includegraphics[scale=0.7,angle=0,trim=0 0 0 0]{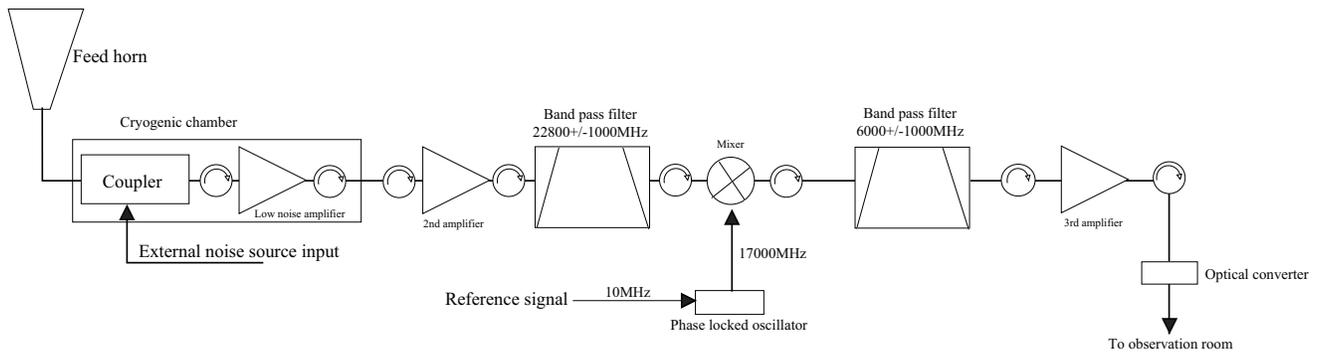}
\caption{Block diagram of the 22 GHz front-end receiver of Kashima 34 m diameter radio telescope. }
\label{fig:kas22g} 
\end{figure}

We conducted observations of water maser sources using the 34-m-diameter radio telescope of the Kashima Space Technology Center, National Institute of Information and Communications
Technology (NICT)\footnote{http://www2.nict.go.jp/aeri/sts/stmg/index\_e.html}, on October 28, 2013, under good weather conditions. We measured  system noise temperature, $T_{\rm sys}$ by the R-Sky method at the zenith at 04:15 (UT). The measured $T_{\rm sys}$ was 146 K, which was the best performance throughout the year. The system equivalent flux density (SEFD) was estimated to be 1025 Jy. The beam size and aperture efficiency were 0.023$^{\circ}$ and 40 $\pm$ 2 \% at 22 GHz, respectively. The pointing accuracy was 4.38/1000 deg for the azimuth angle and 3.94/1000 deg for the elevation angle. The polarization was fixed to left-handed circular polarization.
Three strong water maser sources, W3 (H$_2$O), W49N, and W75N were observed at 22 GHz
for 10 min each. The received signals in the radio frequency range of  22012 -- 22524 MHz (see figure \ref{fig:kas22g} for a block diagram of the 22GHz receiver) were converted
to intermediate-frequency signals of 512 -- 1024 MHz
 via two-step frequency conversion and recorded with two-bit quantization at a sampling rate of
 1024 MHz using an ADS3000+ sampler (\citealt{2010ivs..conf..378T}). 
 
 \section{Results }
\begin{figure}[htbp]
 \includegraphics[scale=2,clip]{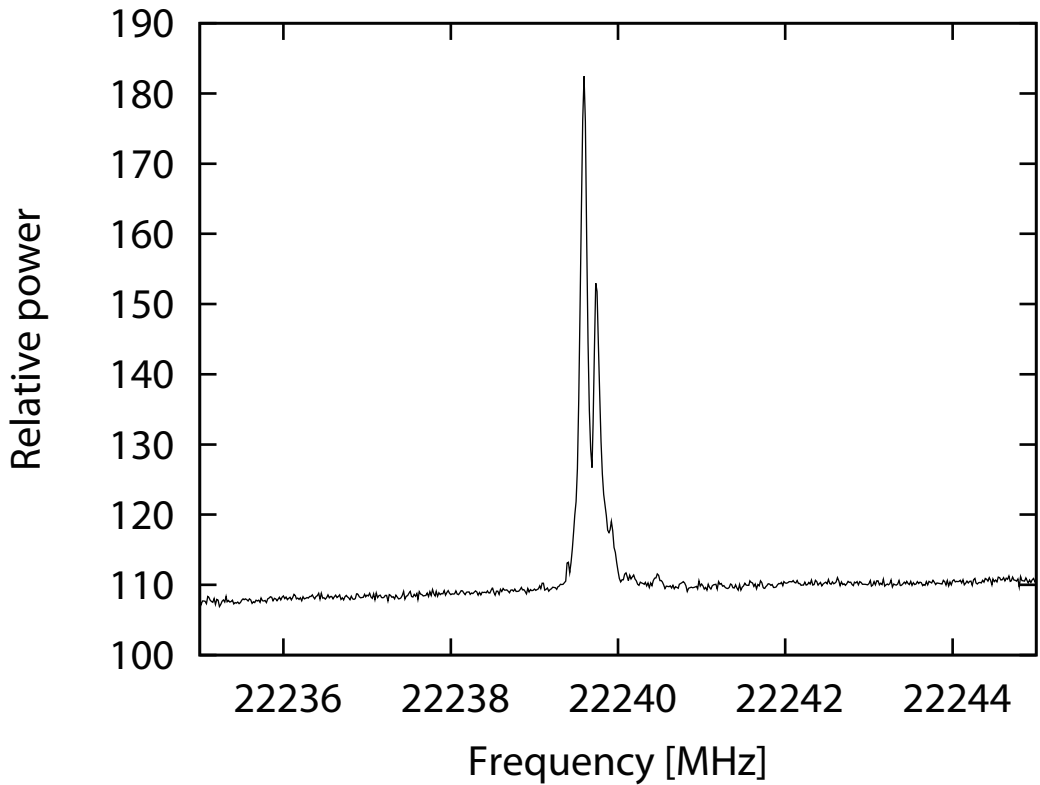}
 \caption{Total-power spectrum of the W3 (H$_2$O) water maser. The spectrum was obtained by general spectroscopy, observed on 28 October 2013 8:49 UT.
 The figure shows the total-power spectrum of the water maser as above with
10 s integration. The spectral resolution is 15.625 kHz,
corresponding to a radial velocity resolution of 0.213 km~s$^{-1}$. 
 The full width at half power (FWHP) of the strongest maser peak is about 96 kHz.}
 \label{fig.w3oh.spec}
\end{figure}

\begin{figure}[htbp]
\includegraphics[scale=2,trim=0 0 0 0]{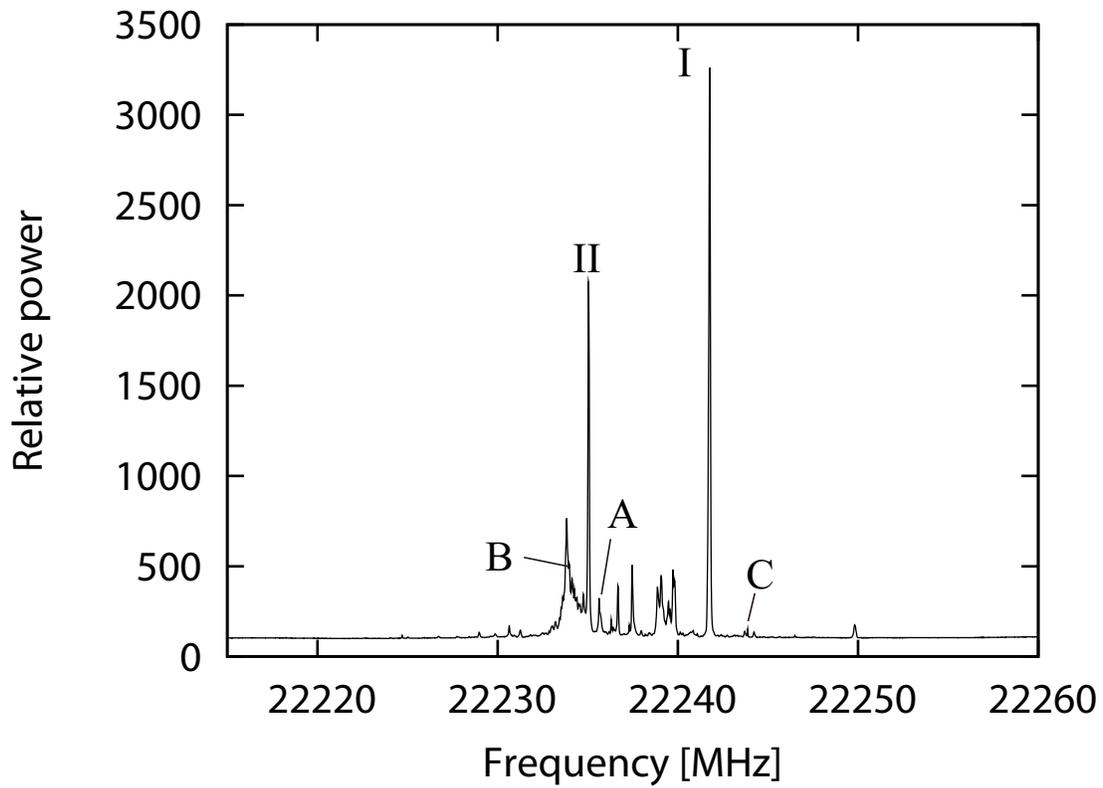}
 \caption{Same as figure \ref{fig.w3oh.spec} but for the W49N water maser, observed on 28 October 2013 8:09 UT.
 W49N is composed of many radial velocity components. The FWHPs of the two strong
 peaks are 101 kHz at 22241.8 MHz (label I) and 72 kHz at 22235.0 MHz (label II). Labels (e.g. A, B, and C) denote the spectrual peaks with the three longest coherence times. Also see figure \ref{fig.cohe.w49n} for comparison. } 
\label{fig.w49n.spec}
\end{figure}

\begin{figure}[htbp]
\includegraphics[scale=2,trim=0 0 0 0]{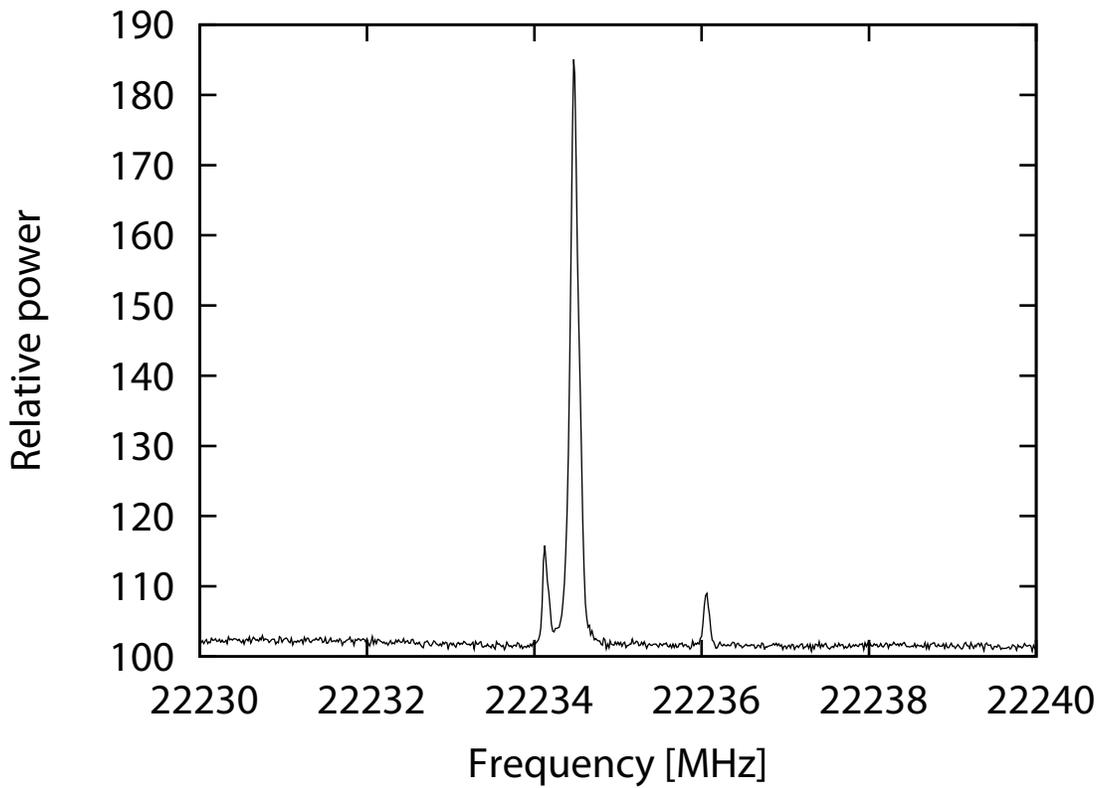}
 \caption{Same as figure \ref{fig.w3oh.spec} but for the W75N water maser, observed on 28 October 2013 9:18 UT. } 
\label{fig.w75n.spec}
\end{figure}

\begin{figure}[htbp]
\includegraphics[scale=.8,angle=0,trim=0 0 0 0]{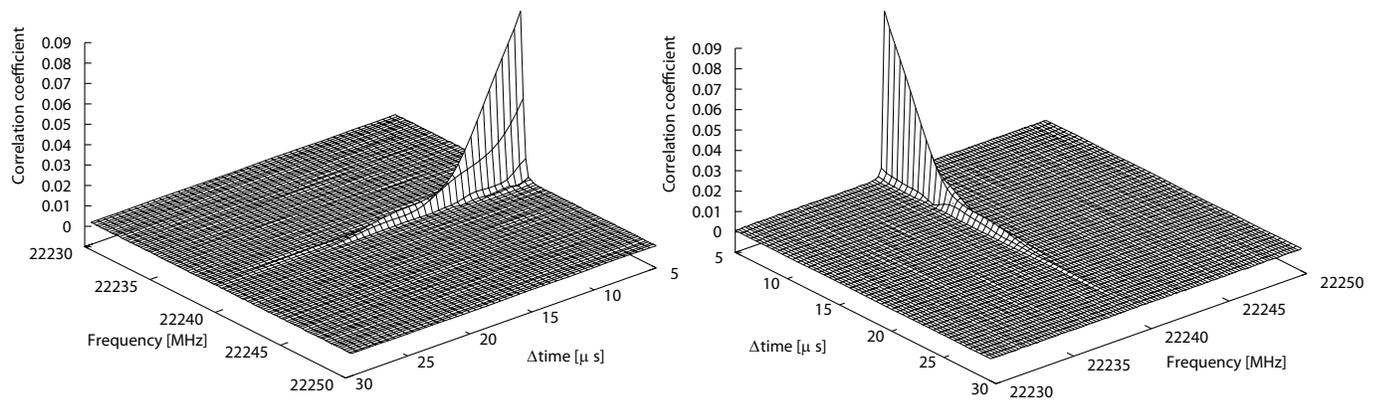}
 \caption{3-D coherence structure of W3 (H$_2$O) water maser computed by XCS with a minimum time spacing of 500 ns,  
 at frequency resolution of 200 kHz and an averaging
 time of 30 s. The resolution gives a forbidden time lag of 5 $\mu$s.
 } 
\label{fig.w3oh.xcs}
\end{figure}

\begin{figure}[htbp]
\includegraphics[scale=.8,angle=0,trim=0 0 0 0]{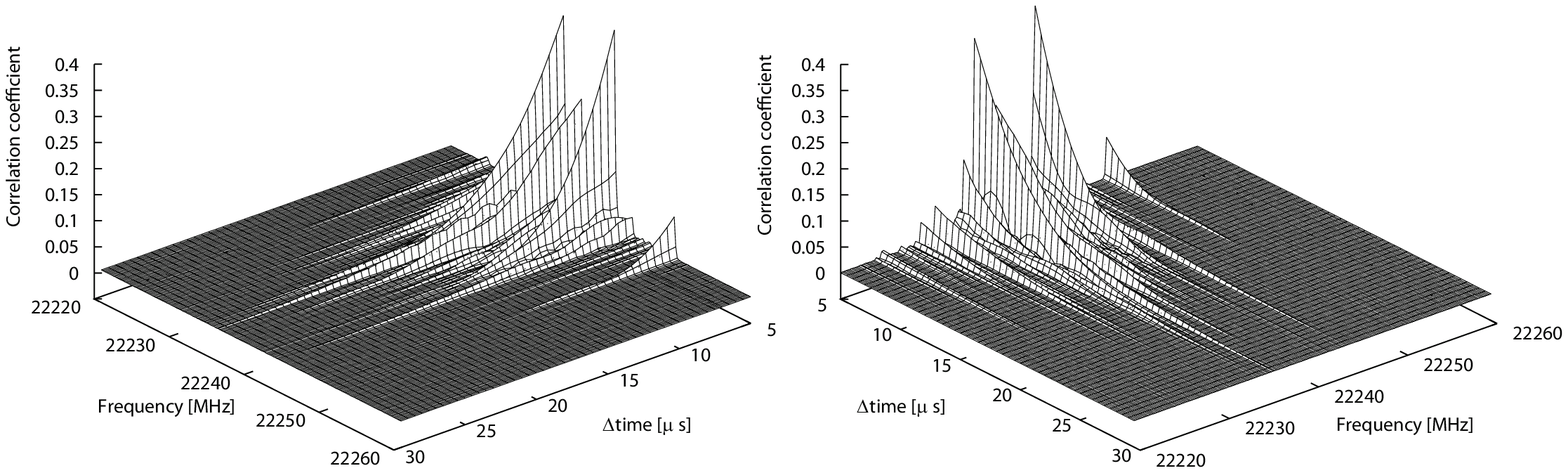}
 \caption{Same as figure \ref{fig.w3oh.xcs} but for the W49N water maser. }  
\label{fig.w49n.xcs}
\end{figure}

\begin{figure}[htbp]
\includegraphics[scale=.8,angle=0,trim=0 0 0 0]{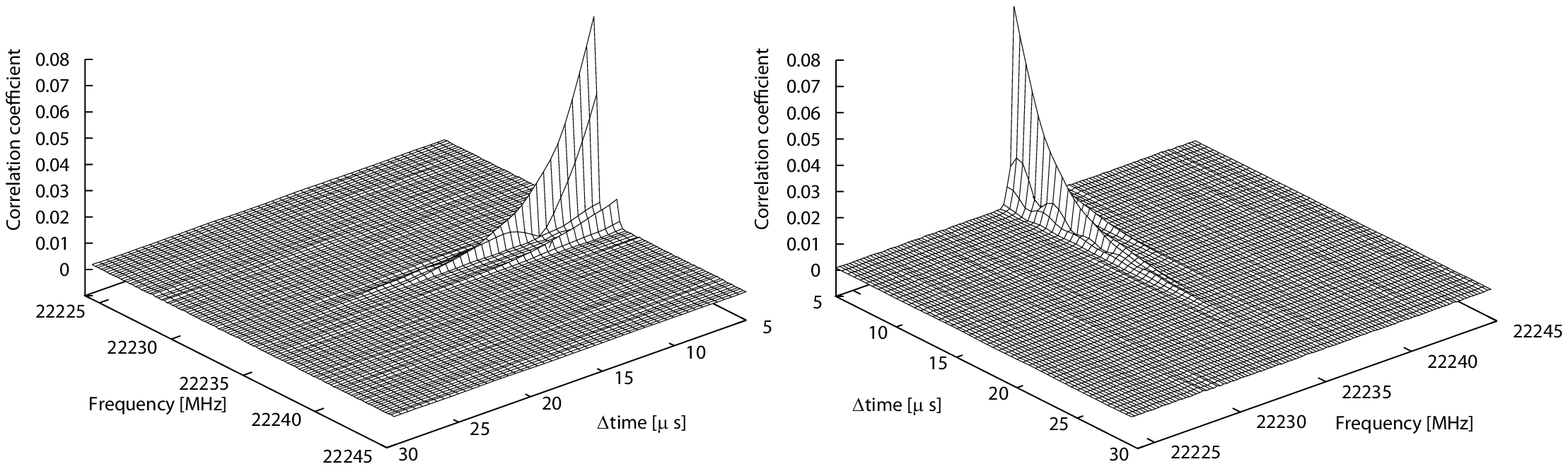}
 \caption{Same as figure \ref{fig.w3oh.xcs} but for the W75N water maser. } 
\label{fig.w75n.xcs}
\end{figure}

 We developed XCS software with C++ and applied XCS to the maser sources. 
 Figures \ref{fig.w3oh.spec} to \ref{fig.w75n.spec} show the total-power spectra of the water masers obtained by general spectrometry with 10 s integration. The spectral resolution is 15.625 kHz,
corresponding to a radial velocity resolution of 0.213 km~s$^{-1}$.  
Figures \ref{fig.w3oh.xcs} to \ref{fig.w75n.xcs} show right and left sides of three-dimensional plots of the coherence
coefficients computed in the XCS processing using equation (\ref{eqn:cohrence-cff}) in a frequency resolution of 200 kHz and an averaging
 time of 30 s. The minimum forbidden time lag was 5 $\mu$s due to the frequency resolution of 200 kHz. Time-series data were shifted by a unit of 500 ns in addition to $\Delta T$(= 5 $\mu$s). 
  The figures \ref{fig.w3oh.xcs} to \ref{fig.w75n.xcs} can be considered to reflect only correlated signals from coherent maser emission, because unwanted signals like system and sky noises will be suppressed by their uncorrelated features.
The most of the 3D coherence structures of the maser sources have clear envelopes. However, some sawtooth shapes against time can be seen in the 3D coherent structures (e.g., W49N at 22234.8 MHz, W75N at 22234.2 MHz and at 22234.6 MHz ). They are produced by the multiple lines within the frequency resolution bin because of the sawtooth shape  and the multiple lines within the frequency resolution bin ($\pm$ 100 kHz) are a Fourier transform pair.

\begin{figure}[htbp]\centering 
\includegraphics[scale=2,angle=0,trim=0 0 0 0]{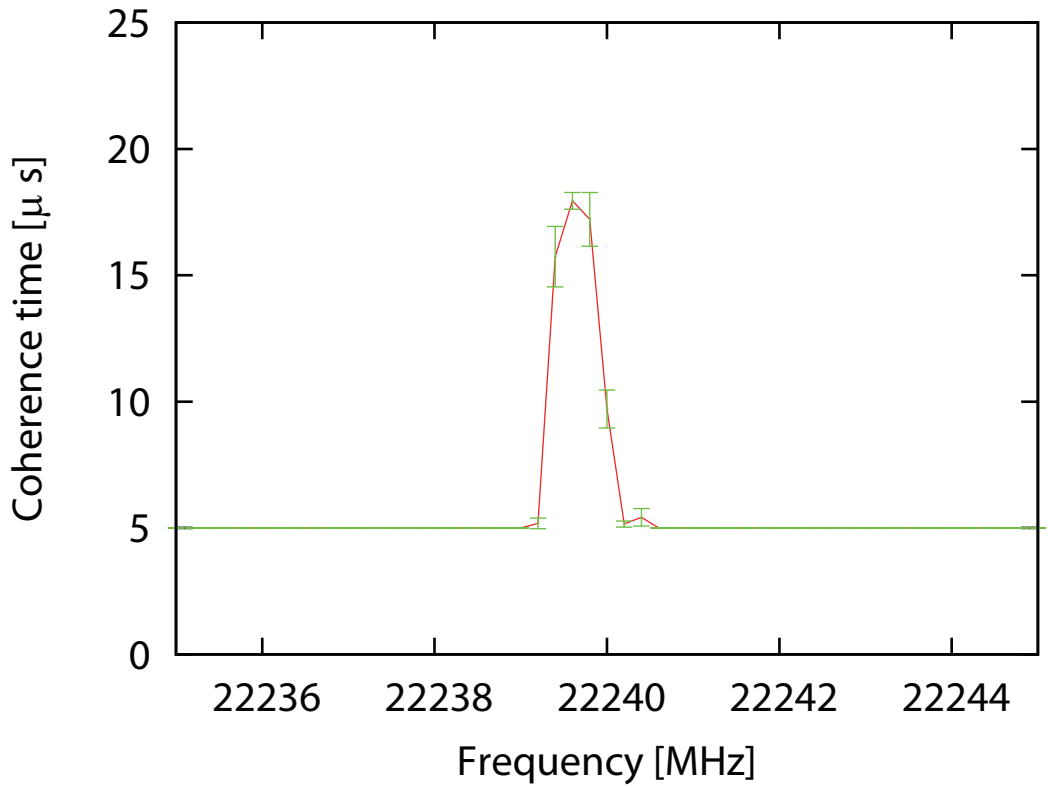}
\caption{Coherence time of the W3 (H$_2$O) water maser obtained from 19 datasets with each 30 s integration data as used to display figure \ref{fig.w3oh.xcs} against frequency.   }
\label{fig.cohe.w3oh} 
\end{figure}

\begin{figure}[htbp]\centering 
\includegraphics[scale=2,angle=0,trim=0 0 0 0]{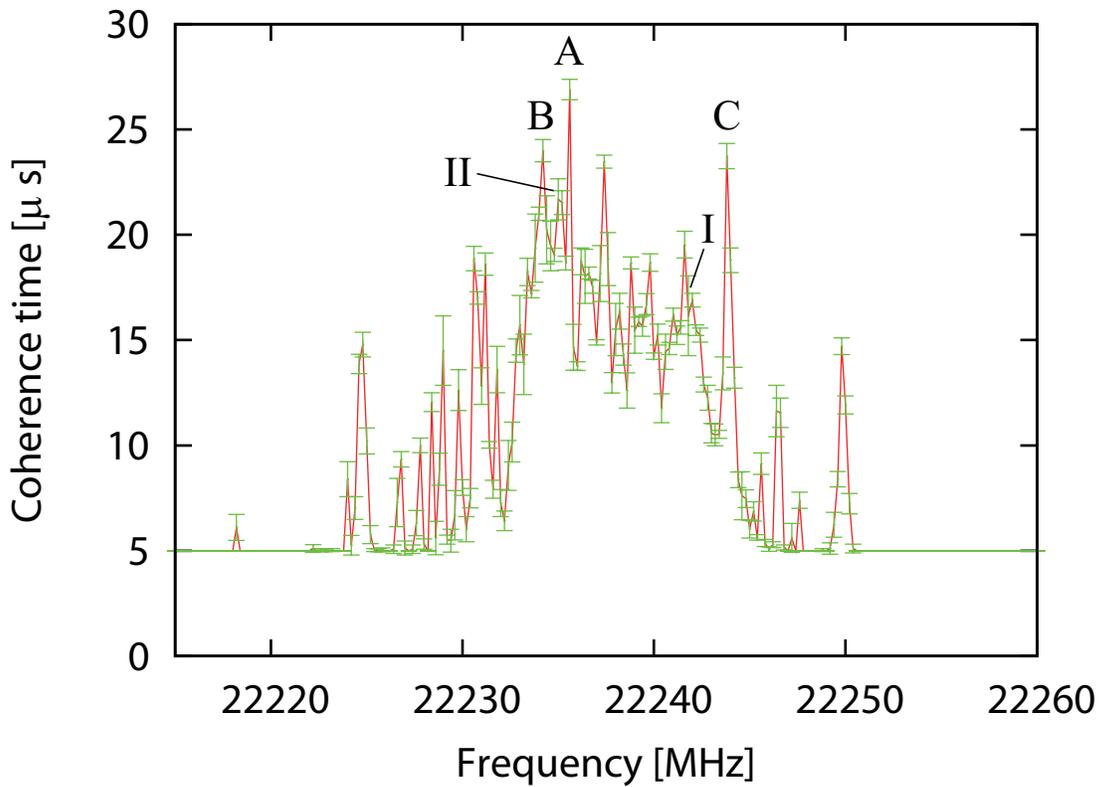}
 \caption{Same as figure \ref{fig.cohe.w3oh} but for the W49N water maser derived from figure \ref{fig.w49n.xcs}. Labels (A, B, C, I, and II) are the same as those figure \ref{fig.w49n.spec}.  }
\label{fig.cohe.w49n}
\end{figure}

\begin{figure}[htbp]\centering 
\includegraphics[scale=2,angle=0,trim=0 0 0 0]{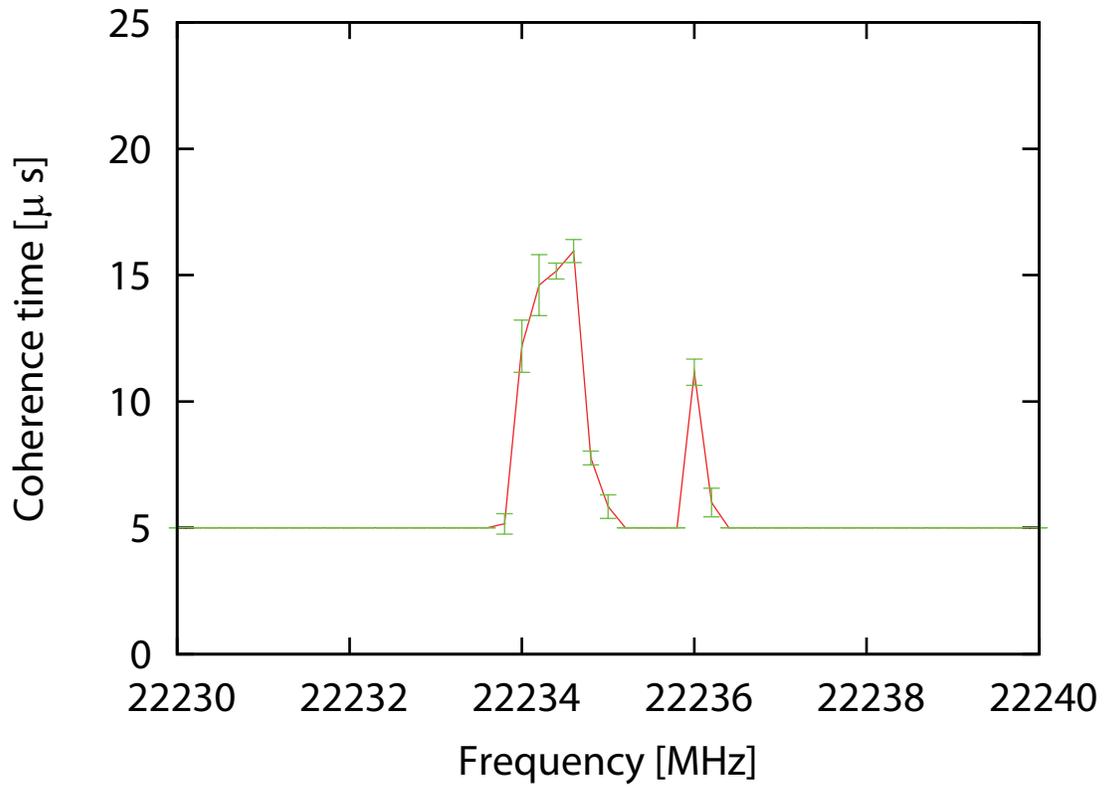}
 \caption{Same as figure \ref{fig.cohe.w3oh} but for the W75N water maser derived from figure \ref{fig.w75n.xcs}  }
\label{fig.cohe.w75n}
\end{figure}

We measured the temporal coherence time at which the maser signals decreased and almost disappeared in the noise in these 3-D coherence structures in the following way.
Firstly, we performed XCS processing for the three masers on our 10 min observation data to obtain 19 datasets with each 30 s integration. Then we obtained the mean correlation amplitude and the noise level, in which any maser signals did not included, at every lag time from the 19 datasets.
 Second, we searched time shifts, where correlation amplitudes become larger than 0.003 corresponding to 6.6 times of the noise level, and defined their first positions as coherence time. This time search was done in the direction from 30 $\mu$s to 5 $\mu$s taking account of the sawtooth features.
 We repeated the way of evaluating to the 19 datasets. Finally, we obtained the mean coherence time and its standard deviation for the three masers shown in figure \ref{fig.cohe.w3oh} to \ref{fig.cohe.w75n}. 
The measured maximum temporal coherence time of three masers are 17.95 $\pm$ 0.33 $\mu$s at 22239.6 MHz, 26.89 $\pm$ 0.49 $\mu$s at 22235.6 MHz and 15.95 $\pm$ 0.46 $\mu$s at 22234.4 MHz for W3 (H$_2$O), W49N and W75N, respectively. The derived coherence time was affected by the noise level, the uncertainty in the derived value due to the noise may be less than 0.4 $\mu$s. Owing to this short coherence time, we required a short Fourier transform length. 
 Interestingly, the strongest maser line at 22241.8 MHz in the W49N in figure  \ref{fig.w49n.spec} does not have a longest coherence time and has a coherence time of 16.16 $\pm$ 1.90 $\mu$s in figure  \ref{fig.cohe.w49n} (label I). On the other hand, the maser line at 22234.2 MHz (label B) in the W49N in figure \ref{fig.w49n.spec} is buried in the complex shapes. However, the coherence time at 22234.2 MHz in the W49N in figure \ref{fig.cohe.w49n} appears the second longest coherence time of 24.00 $\pm$ 0.53 $\mu$s. Thus, a strong maser signal does not have a long coherence time categorically.

\begin{figure}[htbp]\centering 
\includegraphics[scale=2,angle=0,trim=0 0 0 0]{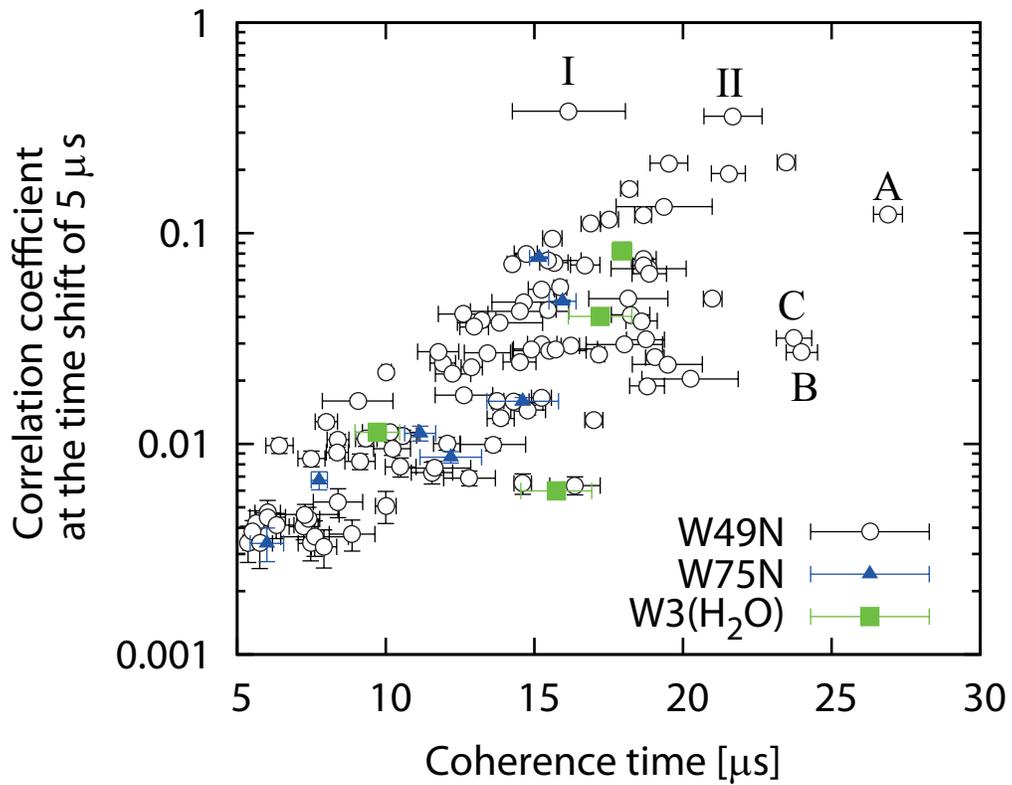}
 \caption{Correlation amplitude at the time shift of 5 $\mu$s against coherence time of W3 (H$_2$O) (green square), W49N (black open circle), and W75 (blue triangle). Labels (A, B, C, I, and II) are the same as those figure \ref{fig.w49n.spec}.}
\label{fig.comp}
\end{figure}

 Figure \ref{fig.comp} shows the coherence time of figure \ref{fig.cohe.w3oh} to \ref{fig.cohe.w75n} against the correlation amplitude at the time shift of 5 $\mu$s of figure \ref{fig.w3oh.xcs} to \ref{fig.w75n.xcs} of three masers.
  We can see that the maximum correlation amplitudes over coherence time are approximately on the line connected between the correlation amplitude of 0.01 at the coherence time of 5 $\mu$s and the point labeled by II.
  According to the maser theory \citep{elitzur},  unsaturated maser exponentially increases its intensity with the path length of the maser spot. Therefore, since observed correlation amplitude of the maser may be directly proportional to the intensity of the maser considered to be a point source, the straight line may suggest a property of the unsaturated maser. Only the strongest maser line of W49N (label I) has a large correlation coefficient over the straight line. The strongest line might be produced by an aggregation of weaker maser lines.   

\section{Discussion \label{sub:dis}} 
\begin{figure}[htbp]\centering 
\includegraphics[scale=.5,angle=0,trim=0 0 0 0]{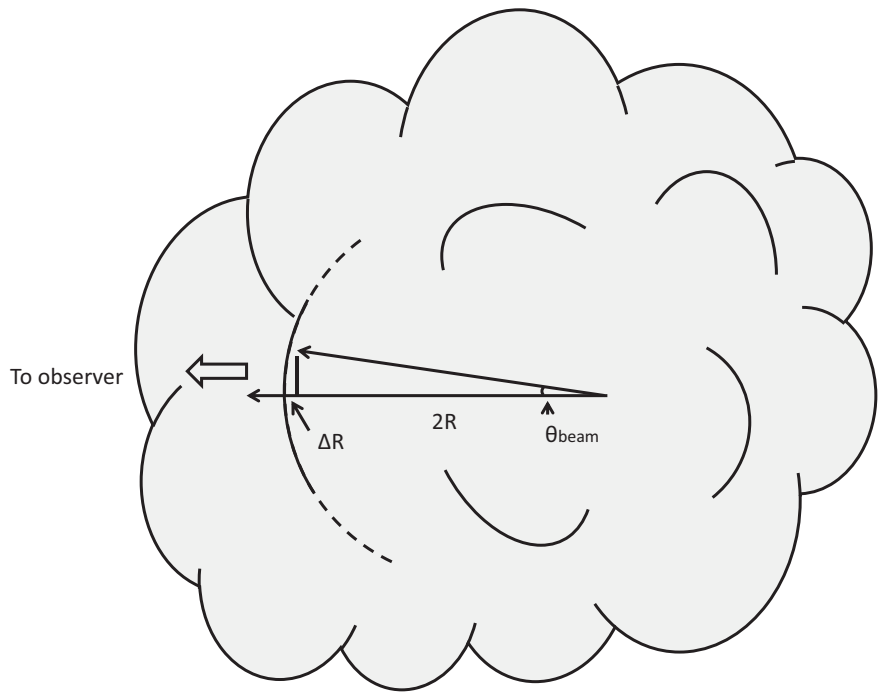}
 \caption{Schematic concept of coherent maser region. }
\label{fig.region}
\end{figure}

The derived coherence time is much shorter than a light travel time of a path length that corresponds to the expected size of a water maser spot ($\sim$ 1 AU, e.g., \citealt{1981ARA&A..19..231R}). This is consistent with the expectation that observed astronomical masers are usually incoherent and coherent maser regions, even if exist, are very tiny. 
Nevertheless, if one supposes such a coherent maser, it may have a uniform physical
condition in order to maintain coherency, namely an equal maser gain per gas volume.
In addition, in a bundle of maser amplification path lengths should be equal within a
certain threshold of coherent maser region. In practice, the coherent region covers a
limited volume of the whole maser region, including an incoherent maser region, in the antenna beam.
In order to explain a very tiny area of the observed coherent maser in the whole maser region, one can suppose a local spherical morphology of the coherent maser region, in which only the maser radiation transferred through the geometrical center of the convexity may have the maximum length of the maser amplification. 
With taking into account a maser beaming angle, $\theta_{beam}$, we consider the difference of maser path lengths between the center and the convex boundary of the coherent maser region as shown in figure \ref{fig.region}. For the maser region with a length $2R$, the difference in the maser path length, $\Delta R$, is derived as a function of $\theta_{beam}$ to be,
\begin{align}
 \Delta R = 2R - 2R \cos \theta_{beam} \\
  \simeq R \theta^{2}_{beam}. \label{eqn:radial}	
\end{align}
 Now this path length difference may give a coherence time $\Delta t_{coh}$ (typically about 20 $\mu$s based on our measurements),
    \begin{equation}
    \Delta t_{coh} \sim \frac{\Delta R}{c}, 
    \end{equation}
    where  $c$ is the speed of light [ms$^{-1}$].
Thus $\theta_{beam}$ is derived to be,
\begin{equation}
\theta_{beam} \simeq (\frac{\Delta R}{R})^\frac{1}{2} . \label{eqn:maserbeamingangle}
	\end{equation}
	If $2R$ = 1 AU (about 500 s for the light speed), we roughly obtain the beaming angle $\theta_{beam}$ and the beaming solid angle $\Omega_{coh}$ to be about $ 	(\frac{20 \times 10^{-6}}{250})^\frac{1}{2} = 2.8 \times 10^{-4}$ rad and  $8.0 \times 10^{-8}$ str, respectively.  
\\

According to \cite{elitzur}, if the brightness temperature $T_b$ of a maser meets the following value, the likelihood of coherent maser is anticipated,
\begin{equation}
  T_b \gg T_0 \frac{4\pi}{\Omega_m}, \label{eqn:tb}
\end{equation}
where $\Omega_m$ is the beaming solid angle ($\approx 10^{-2} - 10^{-4}$) and 
\begin{equation}
  T_0 \equiv \frac{2 \pi h \nu^2\Delta\upsilon}{kcA},
\end{equation}
where $h$ is Planck's constant [Js], $k$ is Boltzmann's constant [JK$^{-1}$], $\nu$ is the observation frequency [Hz], $\Delta\upsilon$ is the bandwidth in velocity units [ms$^{-1}$] and $A$ is the Einstein A-coefficient [s$^{-1}$]. $T_0$ is $3 \times 10^{14}$ K for a 22GHz water vapor maser with a bandwidth of 1 km/s.  Thus, the right side of equation (\ref{eqn:tb}) takes a value of $3.7\times10^{17}$ K  to $3.7\times10^{19}$ K.
On the other hand, we estimate the brightness temperatures $T_b$ of W3 (H$_2$O), W49N and W75N on the basis of our observation whether they exceed the brightness temperatures.  Here, table \ref{tbl:para} shows the estimated parameters, which are calculated for the peak flux densities of three masers and the longest coherence time of W49N. Using the Rayleigh-Jeans approximation, the $T_b$ can be written as
\begin{equation}
  T_b = \frac{c^2}{2k \nu^2}I_{\nu}, \label{eqn:tb2}
\end{equation}
where
\begin{equation}
 I_{\nu} = \frac{S_{\nu}}{\Omega_a},  \label{eqn:iv}
\end{equation}
in which $S_{\nu}$ is the flux density [Js$^{-1}$m$^{-2}$Hz$^{-1}$] and $\Omega_a$ is the solid angle of the maser region.
By scaling the peak amplitudes of the masers in figures \ref{fig.w3oh.spec} to  \ref{fig.w75n.spec} by the estimated SEFD (1025 Jy), the total-power flux densities of the three masers of W3 (H$_2$O), W49N, and W75N are estimated.  If the flux densities of the coherent masers have the correlation coefficient at the time shift of 5 $\mu$s in figures \ref{fig.w3oh.xcs} to \ref{fig.w75n.xcs}, the flux densities of the three masers are obtained assuming that they are coherent masers.
The maser beaming angle assuming 2R $=$ 1 AU are obtained by the measured coherence time of three masers by equation \ref{eqn:maserbeamingangle}. Moreover, 
the maser beam cross-sections are obtained by multiplying the maser beaming angle by the size of 1 AU. Since the distances to W3 (H$_2$O), W49N and W75N are 1.95 $\pm$ 0.04 kpc (\citealt{Xu}), 11.11$^{+0.79}_{-0.69}$  kpc (\citealt{2013ApJ...775...79Z}) and 1.30 $\pm$ 0.07 kpc (\citealt{Rygl}), the antenna beam for the maser beam cross-sections are obtained to be of the order of magnitude $10^{-13}$ rad by dividing by the distance to the masers.    
 
Finally we obtain the brightness temperatures of (8.47 $\pm$ 0.41) $\times10^{18}$ K for peak flux density of W3 (H$_2$O), (8.91 $\pm$ 1.28) $\times10^{20}$ K for the longest coherence time of W49N, (6.21 $\pm$ 1.15) $\times10^{22}$ K for peak flux density of W49N  and (4.77 $\pm$ 0.54) $\times 10^{18}$ K for peak flux density W75N from equations (\ref{eqn:tb2}) and (\ref{eqn:iv}). The brightness temperatures of W3 (H$_2$O) and W75N are comparable to the threshold value of the coherent maser but the that for W49N clearly exceed the threshold of a coherent maser. Consequently, the coherent maser should be observable especially toward W49N if our hypothesis is true.
\\

Note that masers may realistically become observational targets of the Radioastron project (\citealt{kardashev}), which operates a 10 m space radio telescope, 
Spektr-R launched by the Russian Astro Space Center in July 2011, can form an ultimately high angular resolution 
interferometer with ground radio telescopes. In the project, astronomical H$_2$O and OH masers have been 
detected using baselines of up to 10 Earth diameters (\citealt{kardashev2}). 
The angular resolution of 36 microarcseconds yielded in the mapping of 
the W3~IRS5 H$_2$O maser emission, for example, corresponds to a linear size of 10$^{7}$ km at a distance of $\sim$2~kpc. 

On the other hand, we suppose a coherent maser region of 1 AU, which will form a maser beam cross-section of the coherent maser region to be $4.2 \times 10^{4}$ km with a supposed beaming angle of $2.8 \times 10^{-4}$ radian. If a coherent maser region is as long as 240 AU, then the maser beam cross-section of the coherent maser will have a size of 10$^{7}$ km. Although the coherent maser region is likely limited into a tiny fraction of the volume of the whole maser gas clump, such a large coherent maser region will be detectable by the space-ground interferometer.
Therefore, further observations in the Radioastron project to trace the temporal variation of the detected maser emission is crucial for identifying maser emission with much smaller sizes and extremely high brightness temperatures as discussed in this paper. Single-dish observations employing the XCS technique will be useful for finding maser sources exhibiting very bright and small structures with coherent properties.

\section{Summary}
 We have developed a cross-correlation spectrometry (XCS) method.    
 XCS is an extension of the autocorrelation procedure and enhances the
 signal-to-noise ratio for signals with temporal coherence.
 As examples, we applied XCS to the observation data for the water
 masers W3 (H$_2$O), W49N and W75N and derived a typical coherence time of 20 $\mu$s. 
 Moreover, we found a linear relation between the coherence time and the maximal correlation amplitude. The relation may suggest a property of the unsaturated maser. 
 If the coherence time is explained by 
 a deviation of maser amplification path lengths
  in a maser beam angle in a spherical coherent maser region, the beam angle  of the coherent maser emission is estimated to be about $2.8 \times 10^{-4}$ radian. According to the assumed brightness temperature, the coherent maser may be observable, especially toward W49N.
 
\section*{Acknowledgements}
 We thank Dr. Vyacheslav Avdeev, who kindly provided us with the interferometry detection results for the maser sources in the Radioastron project on behalf of the Radioastron data processing department. KT thanks Eiji Kawai and Shingo Hasegawa for careful maintenance of Kashima 34 meter telescope and KT could successfully obtain the initial XCS result by 6.7 GHz methanol data provided by Prof. Kenta Fujisawa of Yamaguchi university. 


\clearpage


\begin{landscape}

\begin{table}[htbp]
\begin{center}
\begin{tabular}{ccccccccccccc} 
\hline
Source & \shortstack{Coherence\\ time} & \shortstack{ Correlation \\coefficient} & \shortstack{Flux\\ density}  & \shortstack{Flux density \\as a coherent maser}  & Frequency & Distance  & \shortstack{Maser beaming\\ angle} & \shortstack{Maser beam \\cross-section} & \shortstack{Antenna\\ beam}  & \shortstack{Brightness \\temperature} \\
 &  [$\mu$s] &  & [Jy] & [Jy] &  [MHz] & [kpc] &  [$10^{-4}$rad] &  [$10^{+7}$m] &  [$10^{-13}$rad] &  [K] &  \\\hline
W3 (H$_2$O)$^1$ & 17.95$\pm$0.33 & 0.0824$\pm$0.0009 &   680$\pm$10 &    56$\pm$1  & 22239.6 & 1.95$\pm$ 0.04$^a$  & 2.68$\pm$0.02 & 4.01$\pm$0.04 & 6.67$\pm$0.15  & (8.47$\pm$0.41)$\times10^{+18}$  \\
W49N$^2$    & 26.89$\pm$0.49 & 0.1230$\pm$0.0006 &  2210$\pm$10 &   272$\pm$2  & 22235.6 & 11.11$^{+0.79}_{-0.69}$$^b$ & 3.28$\pm$0.03 & 4.91$\pm$0.04 & 1.43$\pm$0.10  & (8.91$\pm$1.28)$\times10^{+20}$ \\
W49N$^3$     & 16.16$\pm$1.90 & 0.3802$\pm$0.0007 & 29940$\pm$10 & 11380$\pm$20 & 22241.8& 11.11$^{+0.79}_{-0.69}$$^b$ & 2.55$\pm$0.15 & 3.81$\pm$0.22 & 1.11$\pm$0.10  & (6.21$\pm$1.15)$\times10^{+22}$ \\
W75N$^4$     & 15.95$\pm$0.46 & 0.0765$\pm$0.0009 &   830$\pm$10 &    63$\pm$1  & 22234.4 & 1.30$\pm$ 0.07$^c$   & 2.53$\pm$0.04 & 3.78$\pm$0.05 & 9.43$\pm$0.53  & (4.77$\pm$0.54)$\times10^{+18}$ \\\hline
\end{tabular}
\end{center}
\caption{ Parameters obtained by the XCS. See text in discussion for detailed description.\\
1: Peak flux density in figure \ref{fig.w3oh.spec} \\
2: Longest coherence time labeled by A in figure \ref{fig.cohe.w49n}\\
3: Peak flux density labeled by I in figure \ref{fig.w49n.spec} \\
4: Peak flux density in figure \ref{fig.w75n.spec} \\
a: from \citealt{Xu}, b:  from \citealt{2013ApJ...775...79Z}, c: from \citealt{Rygl} 
}
\label{tbl:para}
\end{table}

\end{landscape}

\end{document}